\documentclass[12pt]{article}
\usepackage{a4}
\usepackage{epsfig}
\newcommand{\beao}{\begin{eqnarray*}}
\newcommand{\eeao}{\end{eqnarray*}}
\newcommand{\be}{\begin{equation}}
\newcommand{\ee}{\end{equation}}
\newcommand{\bea}{\begin{eqnarray}}
\newcommand{\eea}{\end{eqnarray}}
\newcommand{\beq}{\begin{eqnarray}}
\newcommand{\eeq}{\end{eqnarray}}
\newcommand{\nn}{\nonumber}

\newcommand{\la}{\lambda}

\newcommand{\Ref}[1]{(\ref{#1})}

\begin{document}

\begin{center}
{\Large\bf  Effective mass of $W$-boson in a magnetic field }

\vskip 10mm {{\large\bf V.V.
Skalozub}\footnote{skalozubv@daad-alumni.de}\hskip 5mm
\\[8pt]
\it Dnipropetrovsk National University,\\
\it 49010 Dnipropetrovsk, Ukraine}
\end{center}


Simple representation  for   the average value of the $W$-boson
one-loop polarization tensor  in  a  magnetic field $B
= const$, calculated in the  ground state of the tree-level
spectrum, is derived. It corresponds to  Demeur's formula for electron  in
QED. The energy of this state, describing
effective particle mass, is computed by solving the
Schwinger-Dyson equation.  As application, we investigate the effective mass squared at
the threshold of the tree-level instability, $ B \to B_c = m^2/e$,
and   show that it is positive.  In this way the
stability of the $W$-boson spectrum is established.    Some
peculiarities of the results obtained and  other
applications  are discussed.
%
%
\section{Introduction}

Nowadays, physics of charged vector particles  in strong magnetic
fields has obtained new stimulus for  investigation. This
concerns, first of all, $\rho$-meson physics where   the effective
Lagrangian describing electromagnetic interactions has been
derived in different approaches
\cite{Djukanovich,Samsonov,Braguta,IIedditch}. In Refs.
\cite{Chernodub10,Chernodub11} on the base of this Lagrangian the
properties of the $\rho$-meson vacuum in strong magnetic fields of
the order $e B \ge m_\rho^2$, where $m_\rho$ -- particle mass,
were investigated and the superconducting state having a structure
similar to Abrikosov's lattice observed. Such type structure was
derived already  in electroweak theory for the $W$-boson
vacuum \cite{Skalozub86}, \cite{Olesen89}, \cite{SV92}. Other
important reasons  are the  existence   of extremely   strong
magnetic fields in the Universe as well as  in
collisions of beams of protons and heavy ions at modern colliders.
In latter case, they influence characteristics and properties of
particles, in particular, $W$-bosons that is important for various
decay processes which are investigated.

Recently   the ground state projection of the  $SU(2)$
polarization tensor for charged  gluons in Abelian chromomagnetic
field has been  calculated and studied at high temperature in Ref.\cite{Bordag2012}. Simple
expression for this function was derived which just corresponds to
the Demeur formula for electron in magnetic field in QED. The
obtained results (eqs. (17), (22) in Ref. \cite{Bordag2012}) can
be modified to find  the ground state energy for charged massive
vector particle. Of course, a number of other  contributions
has to  be added in  different models.

In what follows, we apply the results of Ref. \cite{Bordag2012}
to calculate the  ground state energy, $\langle  t| \Pi(p_\|, B) |
t\rangle,$ for the $W$-boson, accounting for the one-loop
diagrams. Since on the ground state  $|t \rangle$ the full
$W$-boson polarization tensor is diagonal \cite{VRS83}, it is
possible to write down and solve the Schwinger-Dyson (SD) equation
for this state and in this way obtain a nonperturbative effective
mass $M(B)$ (or effective ground state energy) of the particle.
This mass can be calculated for arbitrary values of the field
strength $B$ (at least numerically) that can be useful for various
applications.

In the present paper, the calculated  expression is
investigated in the limit of $B \to B_c$, where $B_c = m^2/e$ is
the critical magnetic field strength for the $W$-boson tree-level spectrum
\bea\label{spectrum}
&& p^2_{0} = p^2_\| + m^2 + (2 n + 1)e B - 2 e \sigma B,
\nn\\ &&(n =  0, 1,\ldots,\qquad \sigma = 0, \pm 1 )
\eea
in a homogeneous magnetic background, $B = const$, described  by the potential
\be\label{potentialB} A_\mu^{ext.} = B x_1 ~\delta_{\mu 2},
\ee
where $p_\|$ is a momentum component along the field direction,
$e$ -- electric charge, $n$ - Landau level number, $\sigma$ - spin
projection. In \Ref{spectrum}  a tachyon mode is present in the
ground state ($|t\rangle = |n=0,\ \sigma = + 1\rangle$) for the
field strength $B \geq B_c = m^2/e$.
Considering this limit, we show that the  effective energy
calculated from the SD  (or gap) equation remains real  for
realistic mass of Higgs particle. Thus, radiation corrections
act to prevent the instability of the vacuum. Other obvious
applications are in $W$-boson physics for different processes in
strong magnetic fields.

The noted problem has been  investigated already in one-loop order
for the $W$ bosons   in the Georgy-Glashow model of electroweak
interactions  (see review \cite{SV92}). As it was found, the
result depends  on the value of the Higgs boson mass $m_H$. For
heavy Higgs particle, $ K = m_H/m_W \geq 1.2$,  the spectrum
stabilization takes place. For light Higgs particle, $K < 1.2$,
the instability  was found. However, this problem was not
investigated in detail for the standard model. Other note,  the
absence of an adequate  representation for the ground state
projection of the  $W$-boson polarization tensor, similar to
Demeur's formula  \cite{Yildits72}, made investigations of
$W$-bosons bulky and complicated.

In the next section, we calculate the $W$- boson polarization
tensor   and   its mean value in the ground state of the spectrum
\Ref{spectrum}.  In sect. 3, we derive the SD equation for the
ground state projection $\langle t |\Pi(B) |t \rangle $ and
investigate the limit  $B \to B_c$.   General conclusions and
discussion are given in the last section.
\section{$W$- boson  polarization tensor}
In what follows,  we use Euclidean space-time and the
representation of the polarization tensor for  gluons given in
\cite{bord08-77-105013}.  It is reasonable to rewrite color gluon
field, $V^a_\mu$  ($a = 1,2,3 $),  in terms of   charged, $
W_\mu^{\pm} = \frac{1}{\sqrt{2}}(V^1_\mu \pm i V^2_\mu),  $  and
neutral, $V^3_\mu = A^3_\mu$, components. In momentum
representation, the initial expression reads
\bea\label{Pi}
\Pi_{\la\la'}(p)&=&\int\frac{dk}{(2\pi)^4} \
    \left\{\Gamma_{\la\nu\rho}G_{\nu\nu'}(p-k)\Gamma_{\la'\nu'\rho'}G_{\rho\rho'}(k)
\right.\nn \\ &&
    +(p-k)_{\la}G(p-k)k_{\la'}G(k)
\nn \\ && \left.
    +k_{\la}G(p-k)(p-k)_{\la'}G(k)\right\}
\nn\\ [6pt]&&
    +\Pi^{\rm tadpol}_{\la\la'} \ ,
\eea
where the second and third lines result from the ghost contribution
and the tadpole contribution is given by
\be\label{Tp}
   \Pi_{\la\la'}^{tadpol} =   \int\frac{dp}{(2\pi)^4} \ \left\{
   2G_{\la\la'}(p) - \delta_{\la\la'}G_{\rho\rho}(p) - G_{\la\la'}(p)\right\}.
\ee
The contributions of the charged tadpole diagrams  are taken into consideration. Only these tadpoles are relevant to the problem of interest.
The vertex factor,
\be\label{Vf}
    \Gamma_{\la\nu\rho}=
    (k-2p)_\rho  \ \delta_{\la\nu}+\delta_{\rho\nu}(p-2k)_\la+\delta_{\rho\la}(p+k)_\nu \ ,
\ee
completes the description of the vector part of the polarization
tensor. These formulas hold also  in a background field, provided
the corresponding expressions for the propagators are used. We
take a homogeneous magnetic background field in the representation
given in eq. \Ref{potentialB}. In this case the operator
components fulfill the commutation relation
\be\label{com} [p_\mu,p_\nu]=i \,e F_{\mu\nu}~~(F_{12} = B).
\ee
In what follows, where it will be not misleading,  we write $B$
instead $e B$ or even  put $B = 1$,  for short. In above formulas
\Ref{Pi} and \Ref{Tp} we omitted the coupling factors $e^2$. These
factors as well as other   factors proper to different models of
interest can be accounted for in the final expressions. Below, we
will also use the notations $l^2=l_3^2+l_4^2$ and
$h^2=p_1^2+p_2^2$, where we  write $l_3$ and $l_4$ for the momenta
in parallel to the background field $p_\| = l_3$ and imaginary
time, respectively. Other information relevant to the massless
case is given in Refs. \cite{Bordag2012,bord08-77-105013}.

To obtain the results for the electroweak sector, one has to take
into account  the masses of the $W$- , $Z $- and Higgs bosons, and
add the contributions of the latter two particles.

First we incorporate the masses in the representation of the
polarization tensor  as given in eq. (51) of Ref.
\cite{bord08-77-105013}. It results from the proper time
representation of the propagators,
\bea \label{st-repr}
&&G(p - k)= \int\nolimits_0^{\infty} d s\ e^{- s  m^2} e^{- s (p - k)^2},
\nn\\ &&G( k) = \int\nolimits_0^{\infty} d t\ e^{- t  M^2} e^{- t  k^2},
\nn\\ &&G_{\la\la'}(p - k)= \int\nolimits_0^{\infty} d s\ e^{- s  m^2} e^{- s (p - k)^2} E_{\la \la'},
\nn\\ &&E_{\la \la'} = e^{ - 2 i e F s}_{\la\la'},
\eea
for  charged and neutral particles, and integration over $k$ in
eq. \Ref{Pi}. The mass $M$ is, $M = 0, m_Z, m_H$ for photon, $Z$-
and  Higgs boson, correspondingly.

The representation for the $SU(2)$ sector  of the standard model
($W$-bosons, massive ghosts and photons)  is obtained in terms of
the  integral over the parameters $s$ and $t$,
\bea\label{Pi1}
&&\Pi_{\la\la'}=
  \int_0^\infty ds\int_0^\infty dt e^{- s  m^2 } \  {\Theta(s,t)}  \left( \sum_{i,j} {M}^{i,j}_{\la\la'}+
                                 {M}^{\rm gh}_{\la\la'}\right)
\nn\\&&\qquad\qquad
+ \Pi^{\rm tadpol}_{\la\la'}
\eea
with
\be\label{Pitp} \Theta(s,t)=\frac{\exp(-H)}{(4\pi)^2(s+t)\sqrt{\Delta}}.
\ee
Here the following notations are used:
\bea\label{nota}    H&=&\frac{st}{s+t}\, l^2+m(s,t)(2 n + 1) B \,,          \nn\\
                    m(s,t)&=&s+\frac12\ln\frac{\mu_-}{\mu_+}\,,   \nn\\
                    \Delta&=&\mu_- \, \mu_+ ,\nn\\
                    \mu_\pm&=&t+\sinh(s)e^{\pm s},
\eea
which are equivalent to eqs. (23-26) in \cite{bord08-77-105013}.
The sum over $i,j$ in \Ref{Pi1}  follows the subdivision
introduced in \cite{bord08-77-105013} and the functions
${M}^{i,j}_{\la\la'}$ are given by eq. (53) in
\cite{bord08-77-105013}.

Now we take the tachyonic projection of $\Pi_{\la\la'}$, eq.
\Ref{Pi1}. In doing so we note  especially $n =0 $ (for $B = 1)$
and the function $\Theta$ simplifies,
\be\label{Th} \Theta(s,t)_{|h^2=1}=\frac{\exp\left(-\frac{st}{s+t}l^2-s\right)}{(4\pi)^2(s+t)\mu_-}.
\ee
For the projection of the functions ${M}^{i,j}_{\la\la'}$ we use
representation (55) in  \cite{bord08-77-105013}. Calculation of
these terms is given in the appendix of Ref. \cite{Bordag2012}. At
this place we mention that under the tachyonic projection we get
directly a representation suitable for further calculations. The
presence of particle masses is reflected in a simple factor in the
integrand of eq. \Ref{Pi1} and not influenced any computation
procedures applied in the massless case.

Note that the expression \Ref {Pi1} is calculated in the Feynman-Lorentz-t'Hooft  gauge
\be \label{HFg} P_\mu W_\mu^{-} - m \phi^{-} = i C^{-},
\ee
in which the  mass of charged ghost, $C^{\pm}$,  and Goldstone, $\phi^{\pm}$, fields equals to the $W$-boson mass $m$.

Detailed calculations of $\langle t| \Pi | t\rangle $ are given in
Ref. \cite{Bordag2012}  and not modified for $m \not = 0$. Only
the contribution from $M^{33} + M^{gh}$ requires an additional
consideration. As it is shown in Ref. \cite{bord08-77-105013},
eqs. (87) - (89), this part can be written in the form,
\be \label{M33} M^{33 + gh}_{\la\la'} = - \int_0^\infty ds~ dt~ e^{- s \, m^2 }\Bigl( \delta_{\la\la'} \frac{\partial \Theta}{\partial s} + E_{\la\la'}\frac{\partial \Theta}{\partial t} \Bigr),
\ee
where $\Theta(s, t)$ is the function in eq. \Ref{Pitp} and the
matrix $E_{\la\la'} = e^{- 2 i s F}_{\la \la'} $.    These
combining into derivatives allow for carrying out one of the
parameter integrations. Using  $\langle t\mid \delta_{\la\la'}
\mid t \rangle = 1$, $\langle t\mid E_{\la\la'} \mid t \rangle =
e^{2 s}$,
\be\label{46}\Theta(s=0,t)=\frac{1}{t^2},\qquad\Theta(s,t=0)=\frac{1}{s\sinh(s)},
\ee
and integrating by part we get in the projection
\bea\label{M4}
&&\int ds\, dt\ e^{ - s\, m^2} \langle t\mid M^{33 + gh}\,\Theta(s,t) \mid t\rangle =
\nn\\ &&\qquad \frac{1}{(4 \pi)^2} \int\frac{dq}{q}\left(\frac{1}{q}+\frac{e^{- q \, m^2} e^{ 2\, q}}{\sinh(q)}\right)
\nn\\ &&\qquad - \frac{m^2}{(4 \pi)^2}  \int_0^\infty ds~ dt~ e^{- s  m^2 } \frac{e^{- s} \exp\left({ -\frac{s t}{s + t} l^2}\right)}{( s + t ) \mu_{-}},\quad
\eea
where in the last line the function \Ref{Th} is substituted. To
complete  this part, we write down  the remaining (except $M^{33 +
gh}$)  terms coming from the main diagram, \Ref{Pi},
\be\label{M3} \langle t\mid \sum_{ij}\bar{M^{ij}}  \mid t\rangle
        =\frac{4}{\mu_-}+ 4\frac{s + t\, e^{2 s}}{s+t}\ l^2,
\ee
and hence
\bea\label{M7}
&&\langle t\mid \Pi( \sum_{ij}\bar{M^{ij}})\mid t\rangle    =\frac{1}{(4\pi)^2}   \int\frac{ds\,dt}{s+t}\, e^{- s m^2}
\nn\\ &&\qquad \times
     \,\left[\frac{4}{\mu_-}+4\frac{s+t\, e^{2 s}}{s + t}\ l^2 \right]
    \ \frac{\exp\left({-\frac{st}{s+t}l^2-s}\right)}{\mu_-}.
\eea
Here, $\bar{M^{ij}}$ reminds about the omitted terms.
The contributions from the tadpoles, \Ref{Tp},  take the form
\bea\label{Mtp}
&&\langle t\mid\Pi^{\rm tp} \mid t\rangle = - \frac{1}{(4 \pi)^2}
\nn\\ &&\quad\times
\int\frac{dq}{q}\, e^{ - q\, m^2} \left(\frac{2+ \cosh(2q) + 3\sinh(2q)}{ \sinh(q)}\right).
\eea
Then we have to add the contributions coming from charged
Goldstone bosons. As computations shown, one term coincides up to
the sign with the last line in eq. \Ref{M4}, and exactly cancels
in the total, as it should be in the renormalizable gauge
\Ref{HFg}. Other contribution  is the tadpole one coming from the
contact vertex $\sim  e^2  \,W^+_\mu W^-_\mu \phi^+ \phi^-$. This
term up to the factor $- 1/4 $ coincides with the first term in
eq. \Ref{Mtp}. Thus, the expressions \Ref{M4} (except the second
line), \Ref{M7} and \Ref{Mtp}, corrected due to the noted tadpole
contribution, represents the electromagnetic part of the $W$-
boson polarization tensor in the projection to the lower, the
tachyonic state. It  corresponds to the Demeur formula for
electron in magnetic field in QED (see Ref. \cite{Yildits72}, eq.
(59)).

Now, we turn to the contributions of the $Z$- boson sector, calculated in the gauge
\be \label{HFgZ} \partial_\mu Z_\mu - i\, m_Z \phi^{Z} =  C^{Z},
\ee
where $\phi^Z $ and $C^Z$ present the  Goldstone and  ghost fields
having the mass $m_Z$. This part can be expressed by using the
obtained expressions eqs. \Ref{M4}, \Ref{M7}. Actually,  according
to  \Ref{st-repr}, one has to introduce in these formulas the mass
factor $e^{- t \,m_Z^2}$. The contribution of the Golstone field
$\phi^z$ and the term in the  second line of \Ref{M4} are canceled
again. But now, after integration by part over the parameter $t$
in eq.  \Ref{M33}, new  term  appears. The sum of contributions
from Goldstones and $M^{33 + gh}_Z$ looks as follows,
\bea\label{M4Z}
&&\int ds\, dt\ e^{ - s\, m^2 - t\, m_Z^2} \langle
t\mid M^{33 + gh}_{Z.Gb}\,\Theta(s,t) \mid t\rangle
\nn\\ &&\quad=
\frac{1}{(4 \pi)^2} \int\frac{dq}{q}\left(\frac{e^{- q\, m^2_Z}}{q}+\frac{e^{- q \, m^2} e^{ 2\, q}}{\sinh(q)}\right)
\nn\\ &&\quad\quad
- \frac{m^2_Z}{(4 \pi)^2}  \int_0^\infty \frac{d s d t~  \exp\left({ -\frac{s t}{s + t}\, l^2}\right) e^{ s}}{( s + t )
\mu_{-}} e^{- s \, m^2 -t\,m^2_Z} ,\qquad \eea
where  the factor $e^{2 \,s}$ in the second line appears from $
E_{\la\la'}$ in eq. \Ref{M33}. Thus,  the contribution from the
$Z$- sector is given by eq. \Ref{M4Z} and eq. \Ref{M7} with
additional factor $e^{- t \, m_Z^2}$ in the integrand.

Then, we restore the couplings and the dimensionality in the
obtained expressions. Remind that  in actual   calculations  we
put  $e B = 1$  and  therefore  the proper-time parameters $s, \,
t, \,q$ became dimensionless. In fact, this means that we measure
them, as well  as the masses, in units of $e B$. Thus, to recover
the dimensionality one has to substitute  $M^2 \to M^2/ (e\,B),
~l^2 \to l^2/ (e\,B)$ and extra total factor $(e \,B)$ coming from
the $\Theta (s,t)$ in eqs. \Ref{Pitp}, \Ref{Th}. For the
electromagnetic sector, we have to introduce the factor $e^2$ in
eqs. \Ref{M4}, \Ref{M7} and the factor $g^2 = e^2/\sin^2 \theta$
for the tadpole contributions eq. \Ref{Mtp}. For the $Z$-boson
sector, the overall factor in eqs. \Ref{M4}, \Ref{M7} is $e^2
\cot^2 \theta$. Here $\theta$ is the Weinberg angle.

The contribution of the Higgs boson sector is given by two diagrams and reads,
\bea\label{PiH}
&&\Pi_{\la\la'}^H(p)= \int\frac{dk}{(2\pi)^4}\left\{\rule{0pt}{14pt}\right.
\nn\\ &&\quad
   (2 k - p)_\la G(p - k, m^2) (2 k - p)_{\la'} G(k, m^2_H)
\nn \\ &&\quad
    + 4\,m^2 \, G_{\la\la'}(p - k, m^2)~ G(k, m^2_H)\left.\rule{0pt}{14pt}\right\} ,
\eea
where we marked the mass of the particle. Again, we have to use
the representation \Ref{st-repr} and then integrate over $k$. In
the ground state projection, the first  line simplifies
considerably because the condition $p_{\la'} \, \mid t
\rangle_{\la'} = 0$  holds. So, only the term $ 4 \,k_\la \,
k_{\la'}$ contributes. The corresponding term up to a factor
coincides with the term $\langle t \mid M^{11} \mid t \rangle $,
eq. (51) in Ref. \cite{Bordag2012}. The second line equals just to
$\langle t \mid E_{\la \la'} \mid t \rangle \Theta(s,t)_{h^2 = 1}$
(see eq. \Ref{Th}).

Thus, for the scalar sector we obtain,
\bea\label{PiH1}
&&\langle t \mid \Pi^H(p) \mid t \rangle =\frac{g^2}{(4 \pi)^2}
    \left\{\rule{0pt}{20pt}\right.
\nn\\ &&\quad  e \,B \int\limits_0^{\infty} \frac{d s d t ~e^{- \frac{ s t }{s + t} \frac{l^2}{e B}} }{(s + t) \mu_-^2}
\exp\left(\!-\!\left[s \left(\frac{m^2}{e B} + 1\right) + \frac{t \,m_H^2}{e B}\right]\right)
\nn \\ &&
    + \, m^2 \int\limits_0^{\infty} \frac{d s d t ~e^{- \frac{ s t }{s + t} \frac{l^2}{e B}}}{(s + t) \mu_-}
\exp\left(\!-\!\left[s \left(\frac{m^2}{e B} - 1\right) + \frac{t \,m_H^2}{e B}\right]\right)
\left.\rule{0pt}{20pt}\right\} ,
\nn\\&&
\eea
where all the necessary factors are substituted.  The expressions
\Ref{M4}, \Ref{M7}, \Ref{Mtp}, \Ref{M4Z} with corresponding
factors  and eq. \Ref{PiH1} give the non-renormalized mean value
of the $W$- boson polarization tensor in the ground state $\mid t
\rangle$ in the standard model. Its renormalization is fulfilled
in a usual way by subtracting of the terms
\bea \label{ct} c.t._1&&= \langle t \mid \Pi(p^2, B, m^2) _{|_ B= 0} \mid t \rangle, \nn \\
c.t._2&&=\langle t \mid \frac{\partial \Pi(p^2, B, m^2)}{\partial
p^2}(l^2 = e B - m^2) _{|_B= 0} \mid t \rangle \eea on the mass
shell of the spectrum \Ref{spectrum} in the ground state $ l^2 = e
B - m^2$ (see Refs. \cite{SV92,Bordag2012} for details).  These
counter terms are divergent at the lower limit $s, t = 0$.

On the base of these formulas, different kind studies  can be
carried out.  In the next section, we investigate  the ground state
energy in the limit of $ B \to B_c$. For this case, the main
contributions come from the upper limit of integrations over the
proper time parameters. So, the renormalization is not important.
\section{Ground state energy at  $B \sim B_c$}
Let us consider the limit of the field strength   $B \to m^2/e$
for calculated expressions.  In this case,   a number of terms is
divergent at the upper limit of integration because of the
smallness of the "effective tree-level mass", $\Delta = m^2 - e
B$, which  enters the cutting factor $e^{- s \Delta} $ going to
unit for $\Delta \to 0$,  and integrals diverge.   These are the last term in the first line of eq. \Ref{M4} and
similar term in eq. \Ref{M4Z}, two last terms in eq. \Ref{Mtp} and
the terms in the second lines in eqs. \Ref{M4Z} and \Ref{PiH1}. The sum of
calculated diagrams obtains the overall factor, $g^2$,  of
$SU(2)_w$ gauge group, and the mass $m_Z$ has to be substituted by
$m$,  due to the relation $e^2 = g^2 \sin^2 \theta$. They   give dominant contributions    and should be accounted
for.

As a result, when all the relevant terms are gathered together, two types of integrals contribute in this limit,
\be \label{Egr} \epsilon^2_t  = \langle t \mid \Pi \mid t \rangle = \frac{g^2}{(4\pi)^2}(I^{(1)} + I^{(2)}).
\ee
First is one parametric,
\be \label{I1} I^{(1)} = - 2 e B \int\limits_c^{\infty} \frac{d q}{q} ~\exp\left[{-q\left(\frac{m^2}{e B}- 1\right)}\right], \ee
where $c $  is a constant of order 1. The second integral is two parametric,
\bea \label{I2}
 I^{(2)} &=& m^2  \int\limits_0^{\infty}d s  d t \frac{\exp\left[{-\left(\frac{m^2}{e B} - 1\right)\frac{s^2}{s + t} }\right]}{(s + t) \mu_-}
\nn\\ && \times
\Bigl( e^{-t \, m^2_H/(e B) } - e^{- t \, m^2_Z/(e B) } \Bigr).
\eea
In the last expression, we used the relation $l^2 = e B - m^2$.
Both of these integrals can be easily estimated.  The first is
negative and equals,
\be \label{I11}
\left.I^{(1)}(B)\right|_{B \to B_0}  = - 2 e B \log \left(\frac{1}{m^2/(e B) - 1}\right) + O(1). \ee
The sign of $I_2$ depends on the relation  between the  masses
$m_H$ and $m_Z$. If $m_H \ge m_Z$, the  second term in eq.
\Ref{I2} is dominant and integral is negative. Otherwise, it is
positive. In the special case, $m_H = m_Z$, $I_2 = 0$.  We get for
the leading term,
\bea \label{I21} &&\left.I^{(2)}(B)\right|_{B \to B_0} = m^2 \log
\left(\frac{1}{m^2/(e B) - 1}\right) \nn\\ &&\qquad\times \left[
\log \left(\frac{2 m^2 + m_z^2}{ 2 m^2 + m_H^2}\right) +
\log\frac{m_Z^2}{m_H^2}\right] + O(1). \eea
Thus, in the standard model the radiation correction to the ground
state energy at the  threshold of instability, $B = B_c$, is
negative for realistic values of the masses $m_H > m_Z$.  Note
that in the Georgy-Glashow model $Z$- boson absences and $I_2 >
0.$

On the base of these  calculations, we  conclude  that   radiation
corrections act to stabilize the tree-level spectrum
\Ref{spectrum}. Really, if one considers  the pole of the
Schwinger-Dyson operator equation taken in the ground state,  $\langle
t|D |t\rangle^{- 1} = \langle t \mid m^2 + l^2_3  - B  - \Pi(  B,  m^2 ,
\Delta \to 0)\mid  t \rangle $, then the positivity of the
"effective mass squared", $m_{eff.}^2(B) =  m^2 - e B +
\epsilon^2_t $,   follows.

This result can be generalized by solving the gap equation for the effective mass.
 Let us do this, assuming for simplicity that $m_H = m_Z$ when  contribution
$I^{(2)}(B) = 0$.

As it is known \cite{SV92}, the ground state $|t> = |n = 0, \sigma
= + 1>$ is the eigenstate for the full polarization tensor. So,
all the radiation contributions result in   an effective energy or
mass. This is key point in deriving the SD equation  for this
state. In general, the $W$-boson polarization tensor is expressed
through four structures: $l^2 = p^2_4 + p^2_3, p^2_{\perp}= h^2 =
p^2_2 + p^2_1, p p , H = p^2_{\perp} - 2 i e F$.  Among these, the
first and the last operator commutes with all others. For the
ground state $|t \rangle$ in addition the Lorentz condition holds:
$p |t \rangle = 0$ \cite{Bordag2012}, \cite{SV92}. So, the
polarization tensor is diagonal on this state. This important
property can serve as motivation for the choice of the $W$-boson
Green function used in the SD equation.

Accounting for noted above, the mean value for the full operator
$\langle t | G^{- 1} |t \rangle $  can be written in  the form:
\be \label{G}  \langle t | G^{-1} | t  \rangle = l^2 + M^2(B) - e B. \ee
Here, we introduced the  parameter $M^2(B)$ which accounts for all
the contributions to ground state  energy - particle mass $m^2$
and field dependent  radiation corrections giving the effective
mass. This definition of the effective mass is more convenient.
Expression \Ref{G} generalizes the structure of the lower Landau
level of the spectrum \Ref{spectrum}. Taking into account  this
property,  as the exact $W$-boson  operator we choice the
expression $G = (p^2 - 2 i e F + M^2)^{-1} $.   We substitute  it
into  the  operator SD equation
\be \label{SDe} G^{-1}(p^2, M^2, F) = p^2 - 2 i e F + m^2 - \Pi(G (p^2, M^2, F)) \ee
  and calculated the r.h.s. in one-loop order. As the ansatz for the ghost and
 Goldstone fields  we substitute  the expressions used in sect. 2 where we replace the mass $m^2 \to M^2$. This is because  in the gauge  \Ref{HFg}
  used  the mass of all the charged fields   is the same and coinciding with the $W$-boson mass. In this case
  the renormalization can  be done according to eq. \Ref{ct}.

With these entities used, in the limit of $B \to B_c$  the  SD equation  for the $\langle t |G^{- 1} |t \rangle$  transforms into gap equation
\be \label{GE}  M^2 - e B  = m^2 - e B  + 2 (e B)\frac{e^2}{(4 \pi)^2} \log (\frac{1}{M^2/(e B) - 1}).
\ee
Here,  in the r.h.s. the one-loop expression \Ref{I11} is substituted.
This equation can
easily be solved graphically by showing  the r.h.s. and l.h.s. in one plot. The value of $M^2$ can be determined at a given field
strength $B$ as  crossing of both curves. In this way a resummation of infinite series of one-loop diagram is fulfilled. In actual calculations,
it is convenient to measure all the variables in terms of  $m^2$. We denote $x = \frac{M^2}{m^2}, y =\frac{e B}{m^2}$. Then eq.\Ref{GE}
takes the form
\be \label{GE1} x = 1 + \frac{2 y}{(4 \pi)^2} \log\bigl(\frac{1}{x/y - 1}\bigr). \ee
  In Figs.1 and 2 we show the results for some values of $y$. As it is occured, the effective mass $M^2$ is positive even
for field strength $e B$ larger than $m^2$. Thus, the
stabilization of the spectrum happens. This is nonperturbative
result   accounting for the influence  of radiation corrections at
the threshold of instability.
~\\~
\begin{center}
\includegraphics[bb=0 0 360 220
,width=0.5\textwidth]{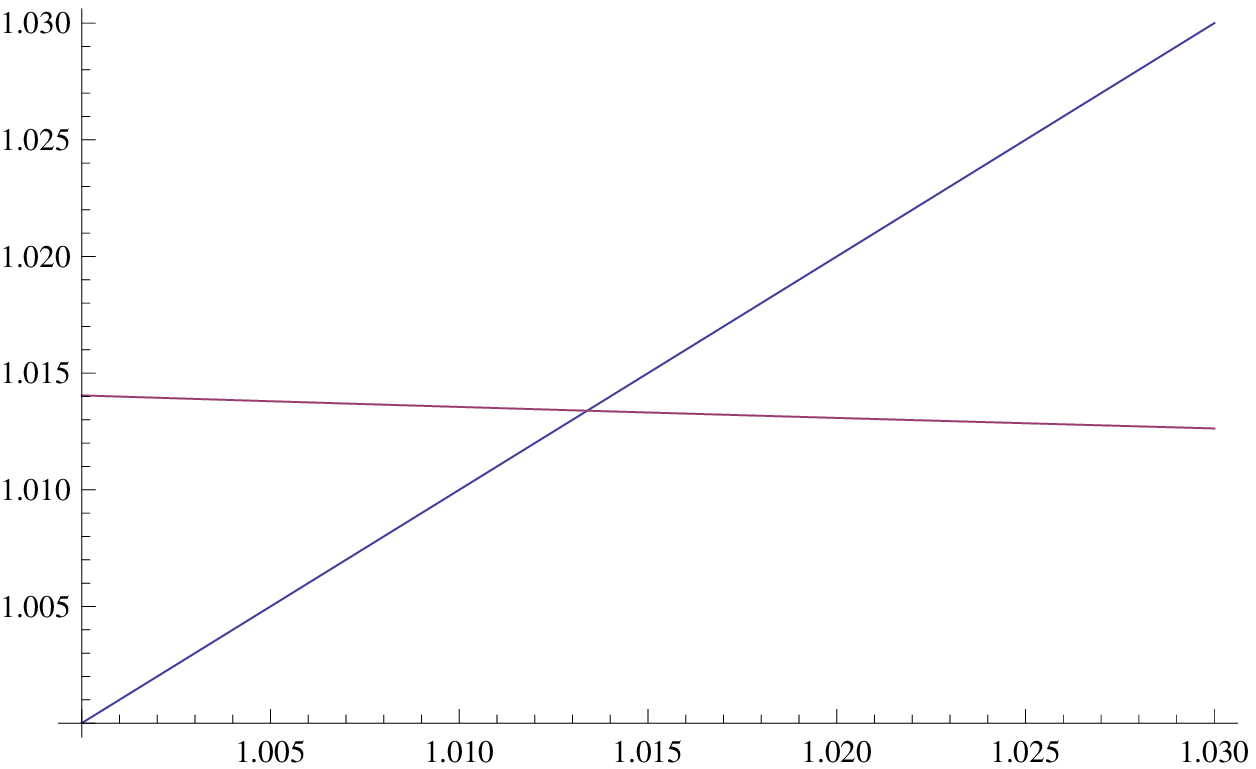}
\end{center}
\mbox{\parbox[t]{0.84\hsize}{\textsf{Fig.1 Effective $W$-boson mass for $y = 0.8$.}}}
~\\~
\begin{center}
\includegraphics[bb=0 0 360 220
,width=0.5\textwidth]{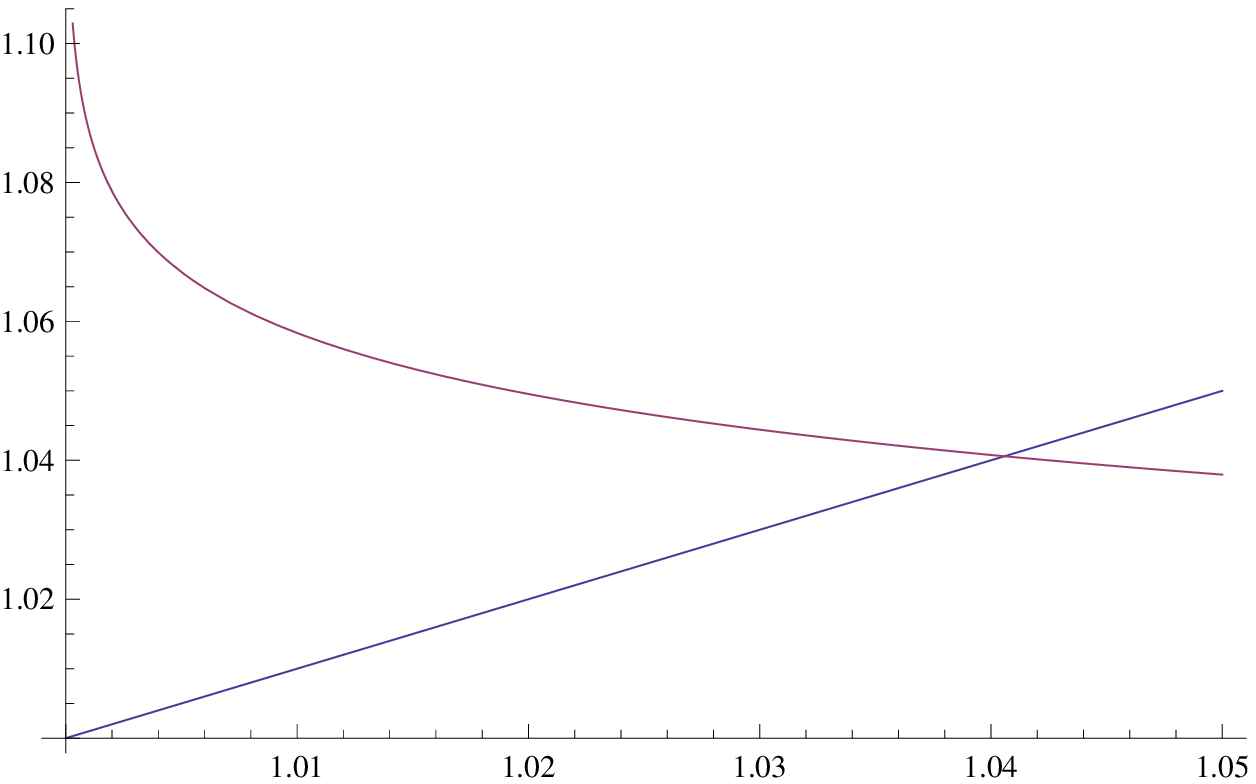}
\end{center}
\mbox{\parbox[t]{0.84\hsize}{\textsf{Fig.2 Effective $W$-boson mass for $y = 1$.}}}
~\\~

To complete this part, we note that the
fermion contribution in one-loop order does note depend on the unstable mode. So, it does
not influence in essential way the effective mass $M^2(B)$.

\section{Discussion and conclusions}
  We derived   simple 
representation for the mean value of the $W$-boson polarization
tensor in external magnetic field calculated in the ground state
of the tree-level spectrum \Ref{spectrum}. It corresponds to the
Demeur expression for electron in magnetic field in QED.  As an
application, we investigated the  behavior of the $W$-boson effective mass
  at the threshold of
instability $B \to B_c$. We found,  the effective mass determined
within the SD equation  is positive, that prevents the vacuum
instability   in strong fields.    This result can be compared
with the one obtained already in one-loop order for the
Georgy-Glashow model \cite{SV92}. In the latter case, however,
there exists the range of not heavy Higgs boson mass for which
radiation corrections shift the threshold of instability to the
weaker than $B_c$ field strengths, and increase instability. In
the standard model such type domain absences  for realistic values
of $m_H > m_Z$.

Note that  in one-loop case  the applicability of
eq.\Ref{I11} is restricted by the condition $\frac{e^2}{(4 \pi)^2}
\log(1/(m^2/(e B) - 1)) << 1$. Therefore, the threshold of
instability is not under control. The solution of the SD equation
is an effective resummation of  infinite series of
 diagrams. As a result, one can investigate not only the fields  $B \sim
B_c$ but also the ones $B \geq B_c$ where the spectrum
stabilization is observed.

The obtained results need in further discussion. In all previous
considerations, the value of the mass $m$ was taken as being
fixed. But this is not the case because  the $W$-boson mass is
determined through the minimum position of the scalar field
effective potential: $m^2 (B)= \frac{1}{2}g^2 \sin^2 \theta
~\delta^2(B)$. The minimum position $\delta(B)$ and the  behavior
of the effective potential depend on the field strength. A
detailed picture  is discussed in review \cite{SV92}. As it 
occurs, in strong fields in the effective potential a second minimum appears. It becomes the
global one and transition to this state happens for the field
strengths close to $B_c$. More details on this point can be found
in \cite{SV92}. So, here we restrict ourselves to this remark and
refer reader to noted paper. For  field strengths not very close
to $B_c$ the initial minimum remains  global and present
consideration is sufficient.

Obtained results on the $W$-boson effective mass  can be applied for arbitrary field strengths $B$.
 In general case, the SD equation in the ground state projection can be solved numerically. It determines
  the nonperturbative effective mass of particle,  that could find applications for
 studying of various processes with $W$-bosons.  For instance, in physics at the LHC where strong magnetic fields
 have to present.

\section*{Acknowledgements}
The author is grateful to Michael Bordag for careful reading  the manuscript, fruitful discussions and suggestions.

\end{document}